\documentstyle[namedreferences]{kluwer}
\input psfig.sty

\begin{opening}
\title{Column Density Distribution of Galactic Ly$\alpha$ Absorption Systems}
\subtitle{(Letter to the Editor)}

\author{Milan M. \surname{\'{C}irkovi\'{c}}\thanks{Affiliated to Astronomical Observatory, 
Volgina 7, 11000 Belgrade, SERBIA}
}
\institute{Dept. of Physics \& Astronomy,\\ 
SUNY at Stony Brook,\\ 
Stony Brook, NY 11794-3800, USA\\
{\tt cirkovic@mail.ess.sunysb.edu}}

\author{Srdjan Samurovi\'c}
\institute{Dipartimento di Astronomia,\\
Universit\`a di Trieste, \\
Via Tiepolo 11, \\
I-34131 Trieste, ITALY\\
{\tt samurovi@newton.daut.univ.trieste.it}}

\date{}
\end{opening}
\begin{document}
\begin{abstract}
A simple consistency argument for hypothesis of the galactic halo origin
of the Ly$\alpha$ forest absorption lines is advanced, based on the recent determination of column-density vs.~impact parameter relation for the
low-redshift gaseous galactic haloes. It is shown that observations of
neutral hydrogen absorption around luminous galaxies are consistent with the index of the power-law column density distribution derived from statistical analysis of large samples of high-redshift Ly$\alpha$ forest lines.  

\classification{PACS}{98.62.Ra, 98.62.Gq, 98.80.Es}

\end{abstract}

\section{Introduction: Column Density Distribution}

One of the principal aims of any model for the origin of Ly$\alpha$ absorption systems in QSO spectra is to reproduce the column density distribution function (henceforth CDDF). As noted by Rauch (1998) in QSO absorption studies, the CDDF occupies the same elevated position as the luminosity function in investigation of galaxy systems and distribution. It is usually assumed that it is expressed as the {\it differential\/} distribution function, i.e.~the number of Ly$\alpha$ absorbing systems per unit redshift path per unit neutral hydrogen column density as a function of the neutral column density $N_{\rm HI}$.

The CDDF is traditionally given in the form (Carswell et al.~1984; Milgrom 1988; Hu et al.~1995)
\begin{equation}
\label{cddf2}
f(N_{\rm HI}) = B N_{\rm HI}^{-\beta},
\end{equation}
where $B$ and $\beta$ are positive constants to be fixed by observations in each particular column density and redshift range. The original result of
Carswell et al.~(1984) was that $B = 1.058 \times 10^{11}$ and $\beta =1.68 \pm 0.10$ in the column density interval $13 < \log N_{\rm HI} < 15$ cm$^{-2}$. Newer measurements from the spectra taken with the Keck HIRES suggest the values high-redshift parameters of the Eq.~(\ref{cddf2}) of (Hu et al.~1995; Kim et al.~1997) 
\begin{equation}
\label{par1}
B = 4.9 \times 10^7,
\end{equation}
and
\begin{equation}
\label{par2}
\beta = 1.46_{-0.09}^{+0.05}.
\end{equation}
Similar results have been obtained by other investigators. For instance, in an excellently performed and a particularly well-written study, Press \& Rybicki (1993) obtain the best-fit result for the index of the CDDF
\begin{equation}
\label{par3}
\beta=1.43 \pm 0.04.  
\end{equation}
These results, it should be noted, are obtained by statistical analysis of high-redshift samples of the Ly$\alpha$ forest. 

\section{Column Density Profile of Galactic Haloes}
With the advent of {\it HST\/}, low-redshift Ly$\alpha$ forest has become available for detailed investigation, and several observational surveys (Spinrad et al.~1993; Lanzetta et al.~1995; Chen et al.~1998) showed that a significant fraction of these absorption lines
arose in extended gaseous haloes of normal luminous galaxies. This confirmed old galactic halo model of Ly$\alpha$ clouds, first put forward by Bahcall \& Spitzer (1969). Moreover, some detailed information on the distribution of neutral gas in such huge galactic haloes of characteristic size $\sim 174\, h^{-1}$ kpc has been obtained. The spatial column density distribution of neutral hydrogen around galaxies in low-redshift absorption-selected sample of Chen et al.~(1998) as a function of galaxy impact parameter can be written as
\begin{equation}
\label{chenmain}
\log \left(  \frac{N_{\rm HI}}{10^{20} \; {\rm cm}^{-2}} \right) =- \alpha_1 \log \left( \frac{\rho}{10 \; {\rm kpc}} \right) + \alpha_2 \log \left( \frac{L_B}{L_{B_\ast}} \right) + \alpha_3,
\end{equation}
where $\rho$ is the absorbing galaxy impact parameter, $L_B$ is its B-band luminosity, and $\alpha_1$, $\alpha_2$ and $\alpha_3$ are constants, which are equal to (with 1$\sigma$ uncertainties), $\alpha_1 = 5.33 \pm 0.50$, $\alpha_2 =2.19 \pm 0.55$, and $\alpha_3= 1.09 \pm 0.90$. This is in agreement with earlier results of Lanzetta et al.~(1995), who obtain for the constant $\alpha_1$ value of 5.3 (no errors quoted). 

\section{Inferred CDDF Index}

For galactic haloes, we can apply the same approach as for minihaloes
(Rees 1988; Milgrom 1988) in attempts to infer the CDDF of a statistically significant sample of absorption systems. When the impact parameter is
between $\rho$ and $\rho + d\rho$, the probability for the column density
to be observed between $N_{\rm HI}$ and $N_{\rm HI}+ dN_{\rm HI}$ can be written as
\begin{equation}
\label{ado}
P(N_{\rm HI}) dN_{\rm HI} \propto \rho d\rho.
\end{equation}
It is natural to assume that the observed column density distribution
is proportional to this probability (Murakami \& Ikeuchi 1990). Therefore, we have
\begin{equation}
\label{location}
\frac{dn}{dN_{\rm HI}} \propto P(N_{\rm HI}) \propto \rho \left( \frac{dN_{\rm HI}}{d\rho} \right)^{-1}.
\end{equation} 
In this manner, one can establish connection between the column density-impact parameter relation and the CDDF index. This is well-known argument, which has been used by Rees (1988), Milgrom (1988) and Murakami \& Ikeuchi (1994) to show plausibility of the minihalo model for origin of the Ly$\alpha$ forest lines. For the general case with the isothermal  distribution of optically thin gas, the minihalo model predicts decrease of the physical H~I density with "minihalo-centric" radius as $n_{\rm HI} \propto r^{-4}$. Resulting theoretical CDDF will exhibit index of (Milgrom 1988)
\begin{equation}
\label{mhalo1}
\beta_{mh}=1.5,
\end{equation} 
which is in rather good agreement with the empirical values in the Eqs.~(\ref{par2}) and (\ref{par3}). 

However, there is no reason not to apply the same argument to the normal galactic haloes as well. It has been used in this manner by Lanzetta \& Bowen (1990) for the Mg II-selected sample of absorbing galaxies, but the predicted value of the index of the Mg II column density distribution (or, in their case, equivalent width distribution) function has been in  violation of the observational data, being $\sim 1.5$ times larger than observed. Consequently, they have not attached much significance to it. Probable reason for this is that the variation of column densities of classical metal-line absorption systems with galactocentric radius is necessary much shallower than for Ly$\alpha$ 
lines {\it originating in the same population of clouds\/}, due to effects of metallicity gradients (cf.~Srianand \& Khare 1993). On the other hand, after Chen et al.~(1998) have established the relationship between H~I column density and galaxy impact parameter, as expressed in the Eq.~(\ref{chenmain}), there is no reason not to compare this piece of observational data specific to galactic gaseous haloes with statistical CDDF in large Ly$\alpha$ forest line samples. Thus, from the Eqs.~(\ref{cddf2}), (\ref{chenmain}) and (\ref{location}),
we infer that the exponent of the power-law CDDF can be written as
\begin{equation}
\label{hole}
\beta = \frac{\alpha_1 +2}{\alpha_1} = 1.38 \pm 0.04,
\end{equation}
which is in good agreement with the results in the Eqs.~(\ref{par2}) and (\ref{par3}).  

\section{Discussion}
The similarity between the results in the Eqs.~(\ref{par2}) and (\ref{hole}) has to be interpreted as a coincidence if we believe that normal galaxies do not constitute significant fraction of the total absorption cross-section of the universe. Conversely, it is conceivable that subpopulation of galactic halo absorption systems retains the same CDDF over a large redshift range.

The fact that galactic halo theory seems to be able to account for the global behavior of the column density distribution function can be regarded as another piece of evidence that a significant part of the entire Ly$\alpha$ absorbing population is truly of galactic origin. This is in agreement with mixed population theories, such as developed recently by Chiba \& Nath (1997), which invoke absorption in both galactic haloes and 
minihaloes. This view is strongly supported by results on the redshift evolution of the Ly$\alpha$ forest, suggesting a slow transition between two distinct populations occured at redshifts $z \sim 1.7$ (Bahcall et al.~1996; Weymann et al.~1998). It may be significant that minihalo models predict systematically higher CDDF index than observations suggest for galactic haloes; thus, any mixture would tend to give intermediate values, as is really the case. This may be also be the explanation for the break in the CDDF power-law detected in some studies (Petitjean et al.~1993; 
Rauch 1998, and references therein). 

\acknowledgements{SS acknowledges the financial support of the University of Trieste.}

\end{document}